\documentclass[aps,prl,twocolumn,superscriptaddress,groupedaddress]{revtex4}  
\usepackage{graphicx}  
\usepackage{dcolumn}   
\usepackage{bm}        
\usepackage{amssymb}   
\usepackage{subfigure}
\usepackage{array}
\usepackage{amsmath}
\usepackage{color}
\def\lt{\lambda_\text{t}}

\begin{document}



\title{Theory of microphase separation in bidisperse chiral membranes}
\date{\today}

\begin{abstract}
We present a Ginzburg-Landau theory of micro phase separation in a  bidisperse chiral membrane consisting  of rods of opposite handendness.  This model system undergoes a phase transition from an equilibrium state where the
two components are completely phase separated to a microphase separated
state composed of domains  of a finite size comparable to the twist penetration depth.  Characterizing the phenomenology using linear
stability analysis and numerical studies, we trace the origin of the discontinuous change in domain size that occurs during this to a competition between the cost of creating an interface and the gain in twist  energy for small  domains in which the twist penetrates deep into the center of the domain.
\end{abstract}
\author{Raunak Sakhardande}
\affiliation{Martin Fisher School of Physics, Brandeis University, Waltham,
MA 02453, USA}

\author{Stefan Stanojeviea}
\affiliation{Martin Fisher School of Physics, Brandeis University, Waltham,
MA 02453, USA}

\author{Arvind Baskaran}
\affiliation{Department of Chemistry and Biochemsitry, University of
Maryland, College Park, MD 20742}

\author{Aparna Baskaran}

\affiliation{Martin Fisher School of Physics, Brandeis University, Waltham,
MA 02453, USA}

\author{Michael F. Hagan}

\affiliation{Martin Fisher School of Physics, Brandeis University, Waltham,
MA 02453, USA}

\author{Bulbul Chakraborty}
\affiliation{Martin Fisher School of Physics, Brandeis University, Waltham,
MA 02453, USA}
\maketitle


\paragraph{Introduction:} When two immiscible fluids are mixed, they typically undergo bulk phase separation. Applications ranging from food science, catalysis, and the function of cell membranes require the arrest of this phase separation to form microstructures. A common pathway to accomplish this is the introduction of a third component, such as a surfactant, that stabilizes interfaces between the two fluids \cite{safran2003statistical}. Here, we theoretically demonstrate a novel mechanism for microphase separation in fluid membranes that is mediated by the chirality of the constituent entities themselves, and hence does not require the introduction of a third component. In addition to identifying a design principle to engineer nano structured materials, this work could shed light on the role of chirality in compositional fluctuations and raft formation in biomembranes \cite{hyman2012beyond,lingwood2010lipid,weis1984two,dietrich2001lipid,simons2010revitalizing,veatch2003separation}.

\begin{figure}[ht]

\begin{center}
    \includegraphics[width=0.5\textwidth]{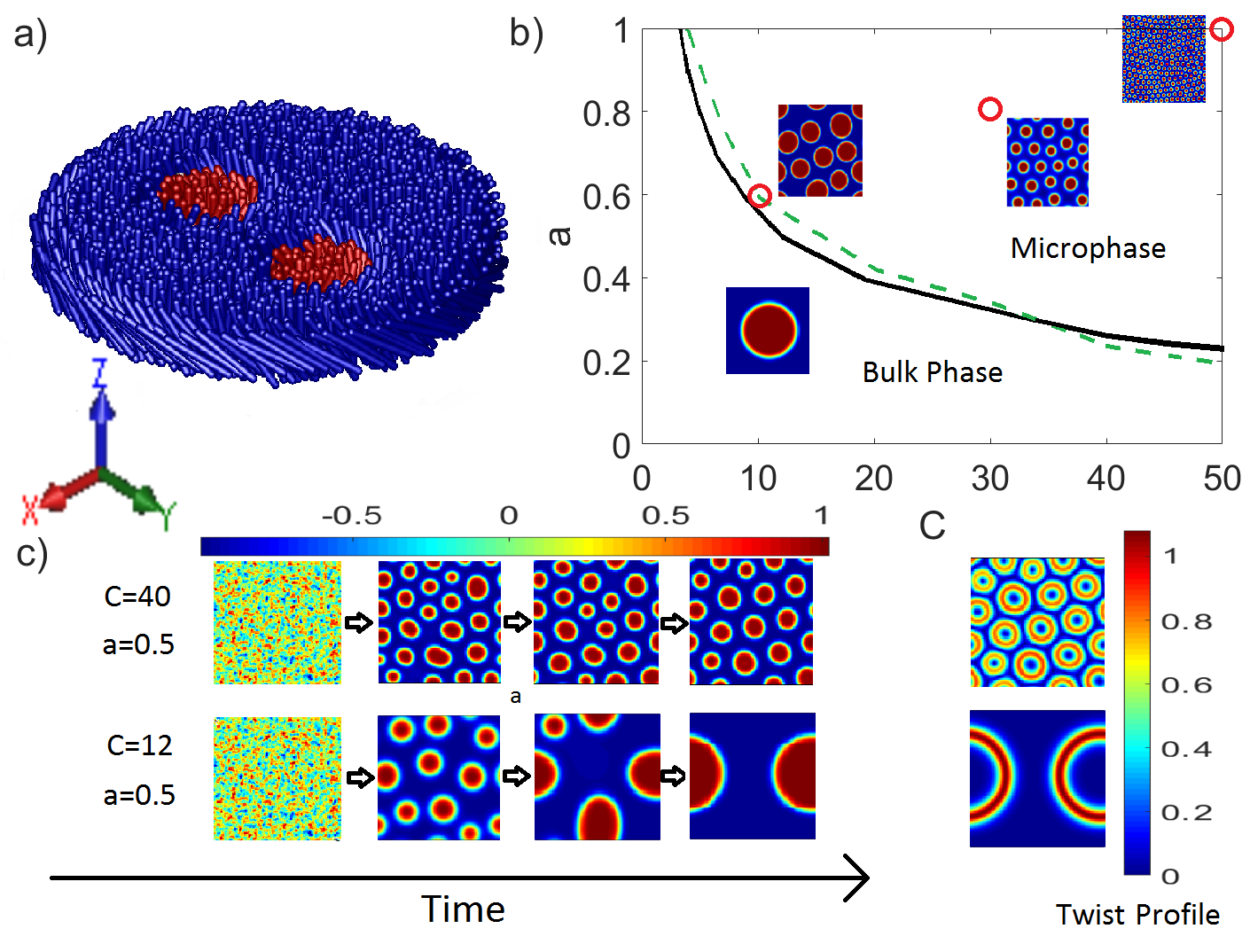}
\end{center}

\caption{a) Schematic of the experimental system. b) The primary results of this work summarized in a phase diagram as a function of the smectic alignment parameter $C$ and the twist-composition coupling parameter $a$. The lines indicate the phase boundary between the microphase separated and bulk phase separated states as obtained from linear stability analysis (black/solid) and numerical integration of Eqs. \ref{dyn1}-\ref{dyn2} (green/dashed). The snapshots show configurations at steady state obtained from numerics at the indicated parameter values ({\color{red} o}). c) Illustration of the evolution to steady state for two parameter sets. The results shown here and in the rest of the paper are for a 60-40 mixture with $\lambda_\psi=0.1$,  and $q_{0}=0.1$}
\label{fig:phase}
\end{figure}

Our theory is motivated by a recently developed colloidal-scale model system of fluid membranes, composed of fd-virus particles \cite{gibaud2012reconfigurable,zakhary2013geometrical,zakhary2014imprintable,barry2009model,sharma2014hierarchical,barry2010entropy}. The system contains two species of virus particles that have opposite chirality and different lengths (Fig \ref{fig:phase}). In the presence of a depletant, they self-assemble into a monolayer membrane that is one rod length thick.  The competition between depletant entropy, mixing entropy of the two species, and molecular packing forces leads to a rich phase behavior within a membrane, including bulk phase separation of the two species, microdomain formation, and homogeneous mixing. In particular, the experiments find that in the regime where a single species forms a macroscopic membrane, limited only by the amount of material, a mixture of two species leads to the formation of circular monodisperse microdomains (rafts) of one species in a background of the other.

In this work, to understand the mechanisms controlling this raft formation, we develop a continuum Ginzburg-Landau theory that captures the physics of chirality and compositional fluctuation in a 2D binary mixture of rods with opposing chiralities. The primary physics that we incorporate into the theory is  a coupling between the twist of the director field and the compositional fluctuations \cite{selinger1993chiral}. By using linear stability analysis and numerical solutions of the time-dependent  Ginzburg-Landau equations, we show that the tendency of the
molecules to twist arrests the phase separation of the two
species, and stabilizes a droplet phase whose phenomenology closely mimics
that observed in experiments. In particular, the theory shows a discontinuous jump in the droplet radius as the system transitions from a microphase separated state to bulk separation, a phenomenon observed in the experiments as well. In contrast, previously studied mechanisms of microphase separation lead to a droplet size that continuously diverges as the system approaches bulk phase separation \cite{elias1997macro,bates2008block,seul1995domain,janssen2007aperiodic}.



\paragraph{Model:}  The Ginzburg-Landau (GL) model involves two fields:
a director field $\hat{\mathbf{n}}\left( \mathbf{r}\right) $ that characterizes the
orientation of the rods with respect to the membrane normal and a scalar
field $\psi \left( \mathbf{r}\right) $,  which characterizes the local
composition of the membrane in terms of the two species. We choose a coordinate system in which the layer normal of the membrane lies along
the $z$ axis and normalize the order parameter $\psi $ such that $\psi =\pm 1
$ correspond to the  homogeneous one-component phases.
The GL functional  is taken to be of the form:
\begin{multline*}
F = \int d^{2}\mathbf{r}\left[ \frac{1}{2}K_{1}( \nabla \cdot \mathbf{
\hat{n}}) ^{2}+\frac{1}{2}K_{2}\left( \mathbf{\hat{n}}\cdot \nabla
\times \mathbf{\hat{n}}-q\left( \psi \right) \right) ^{2} \right.\\ +\left . \frac{1}{2}
K_{3}\left( \mathbf{\hat{n}}\times \nabla \times \mathbf{\hat{n}}\right)
^{2} +\frac{C}{2}\sin ^{2}\theta-\frac{\psi ^{2}}{2}+\frac{\psi ^{4}}{4}+
\frac{\lambda_{\psi} }{2}\left( \nabla \psi \right) ^{2} \right]
\end{multline*}


The physics incorporated in the GL functional can be summarized as follows : i) The first
three terms arise from the Frank elasticity associated
with director distortion, with $K_{1}$, $%
K_{2} $ and $K_{3}$ being the elastic constants associated with splay, twist
and bend respectively \cite{DegenneBook}. The twist term involves a pitch $q(\psi \left( \mathbf{r}\right))$ that encodes the
chirality and hence the associated tendency of the rods to develop a
spontaneous non-zero twist.  In a mixture of left and right handed rods, $q$ is naturally a function of the composition, which introduces a coupling between $\hat{\mathbf{n}}\left( \mathbf{r}\right) $  and $\psi \left( \mathbf{r}\right) $.
ii) The term $\frac{C}{2}\sin ^{2}\theta $ encodes
the fact that the rods in the membrane tend to align with the layer
normal \cite{DegenneBook}, and gives rise to the standard mechanism of twist expulsion
seen in Smectic C systems.
When $\psi =\pm 1$, the terms discussed in (i) and (ii)
reduce to the theoretical description used successfully to describe single
component chiral membranes in earlier works \cite{pelcovits2009twist,kaplan2010theory,kaplan2014colloidal}. iii) The
compositional fluctuations encoded in the field $\psi $ are described by a
standard  $\psi ^{4}$ theory  {\it below} the critical point
that leads to bulk phase separation, with an energetic cost
to forming interfaces controlled by the parameter $\lambda_{\psi} $. Thus, the difference in the length of the rods that leads to phase separation in the experimental system is represented  as an effective interaction, and our 2D model does not include information about the spatial variation of the membrane in the third dimension.

In the following, we work in the single elastic constant approximation of
the Frank elasticity: $K_{1}=K_{2}=K_{3}\equiv K$.   We model the variation of $q$ with composition through a minimal linear  coupling, $q(\psi)=q_{0}+a \psi$, which defines the coupling parameter $a$.
We nondimensionalize the GL functional using the twist penetration depth $\lt\equiv \sqrt{\frac{K}{C}}$  as the characteristic length scale.
Defining dimensionless parameters:  $q_{0}^{\prime
}=\lt q_{0}$, $a^{\prime }=\lt a\left(
1-Ca^{2}\right) ^{1/2},$  $\lambda_{\psi} ^{\prime }=\frac{\lambda_{\psi} /\lambda _{t}^{2}%
}{\left( 1-Ca^{2}\right) }$ ,$\psi ^{\prime }=\frac{\psi }{\left(
1-Ca^{2}\right) ^{1/2}}$ and $C^{\prime }=\frac{C}{\left( 1-Ca^{2}\right)
^{2}}$ the GL functional becomes:%
\begin{align}
F&=\int d^{2}\mathbf{r}^{\prime }\left[f_\text{LC}+f_{\psi}+f_\text{Cross}\right] \nonumber \\
f_\text{LC}&=\frac{C^{\prime }}{2}\left[ \left( \nabla ^{\prime
}\cdot \mathbf{\hat{n}}\right) ^{2}+\left( q_{0}^{\prime 2}-q_{0}^{\prime }
\mathbf{\hat{n}}\cdot \nabla ^{\prime }\times \mathbf{\hat{n}}\right) \right] \nonumber \\
f_{\psi}&=\left[ -\frac{\psi ^{\prime 2}}{2}+\frac{
\psi ^{\prime 4}}{4}+\frac{\lambda_{\psi} ^{\prime }}{2}\left( \nabla ^{\prime
}\psi ^{\prime }\right) ^{2} \right] \nonumber \\
f_\text{Cross}&=\left[ -C^{\prime }a^{\prime }\psi ^{\prime }\left(
\mathbf{\hat{n}}\cdot \nabla ^{\prime }\times \mathbf{\hat{n}}\right)
+C^{\prime }a^{\prime }q_{0}^{\prime }\psi ^{\prime }\right]
\label{eq:GLequation}
\end{align}
This nondimensionalized GL functional is used in all of our subsequent analysis and the $^{\prime }$'s are dropped for compactness of notation.

We model the dynamics by the time-dependent  GL equations with a conserved composition  field
 $\psi $:   $\partial _{t}\psi =\nabla ^{2}\frac{\delta F}{\delta
\psi }$ (Model B dynamics), and a non-conserved director field $\partial _{t}
\mathbf{\hat{n}}=-\left( \mathbf{I}-\mathbf{\hat{n}\hat{n}}\right) \cdot
\frac{\delta F}{\delta \mathbf{\hat{n}}}$ (Model A dynamics)\cite{Hohenberg-Halperin}.
The $\mathbf{\hat{n}}$
dynamics accounts explicitly for the fact that it is a unit vector. The time constants for the relaxation dynamics of $\psi$ and $\mathbf{\hat{n}}$ have been chosen to be same and set equal to 1.  The resulting equations are :
\begin{equation}
\partial _{t}\psi =\nabla ^{2}\left( -\psi +\psi ^{3}-\lambda_{\psi} \nabla
^{2}\psi -Ca\left( \mathbf{\hat{n}}\cdot \nabla \times \mathbf{\hat{n}}%
\right) +Caq_{0}\right)  \label{dyn1}
\end{equation}%
\begin{multline}
 \partial _{t} \mathbf{\hat{n}}=-\left( \mathbf{I}-\mathbf{\hat{n}\hat{n}}\right) \cdot\left(-C\nabla^2\mathbf{\hat{n}}-2Cq\nabla \times \mathbf{\hat{n}}+C\mathbf{\hat{n}} \times \nabla q \right. \\ \left.
+C(n_{x}\hat{x}+n_{y}\hat{y}) \right) \label{dyn2}
\end{multline}

\paragraph{Linear Stability Analysis:} Eqs.(\ref{dyn1}-\ref{dyn2}) admit
homogenous steady states of the form $\psi =\pm 1$ and $\mathbf{\hat{n}}=%
\mathbf{\hat{z}}$.
As a first step in
understanding the dynamics of phase separation, we  analyze the instability of the homogeneous state to small fluctuations of the form $\psi=1+\delta \psi$ and $\mathbf{\hat{n}}=\mathbf{\hat{z}}+\mathbf{\delta n}$. We introduce Fourier transformed variables  $\widetilde{X}\left( \mathbf{k},t\right)
=\int d^{2}\mathbf{r}e^{i\mathbf{k}\cdot \mathbf{r}}X\left( \mathbf{r}%
,t\right) $.
Without loss of generality, we choose a coordinate system in the plane of the
membrane such that the $x$ axis lies along the spatial gradient direction. We find that the longitudinal fluctuations in the director $\delta \widetilde{n}_{x}$ decouple from the other variables (\cite{supinfo}) and we obtain the linearized equations
\begin{eqnarray}
\partial _{t}\left(
\begin{array}{c}
\delta \widetilde{{ \psi} } \\
 \delta \widetilde{ {n}}_{y}
\end{array}
\right) =
 \left(
\begin{array}{ccc}
-2k^{2}-\lambda_{\psi} k^{4} &   ik^{3}Ca  \\
-ikCa & -C-Ck^2
\end{array}%
\right) \left(
\begin{array}{c}
\delta \widetilde{ \psi } \\
\delta \widetilde{ {n}}_{y}
\end{array}%
\right) \label{LSM}
\end{eqnarray}
The homogeneous state  is found to be linearly unstable to modes $k$  that satisfy
\begin{equation}
\lambda_{\psi} k^6-(Ca^2-2-\lambda_{\psi})k^4+2k^2 < 0 ~.
\label{eq:linear}
\end{equation}
\begin{figure}[h!]
\begin{center}
 \includegraphics[width=0.5\textwidth]{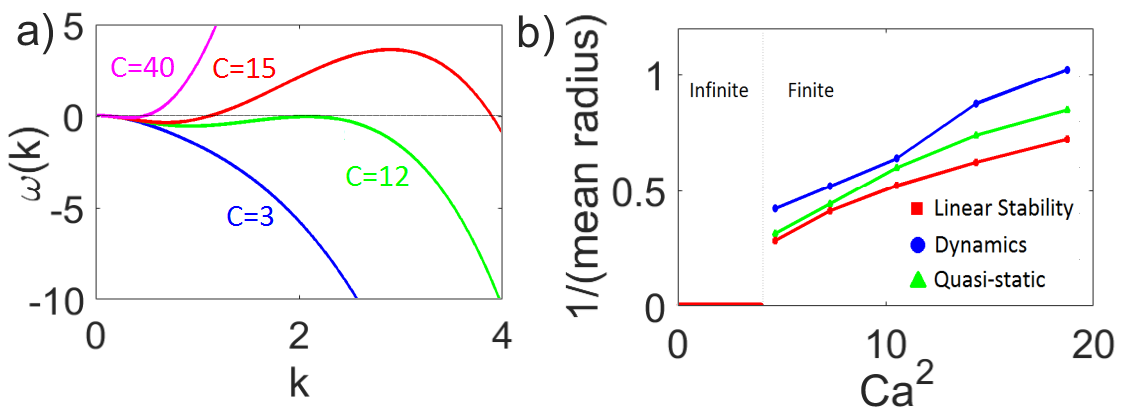}
\end{center}
\caption{(color online) a) The largest eigenvalue $\omega(k)$ of the linear stability matrix in Eq.(\ref{LSM}) as a function of the wavevector $k$ for indicated values of the alignment strength $C$, with $a=0.8$ and $\lambda_{\psi}=0.1.$  b)  Dependence of the optimal domain size on $Ca^2$ obtained by three different analysis methods: the steady-state mean radius of domains obtained by numerical integration, the radius which minimizes the GL free energy (calculated as described in the text), and wavelength corresponding to the fastest-growing mode calculated by linear stability analysis, for $\lambda_{\psi}=0.1$}
\label{fig:dispersion}
\end{figure}

We see from Eq. \ref{eq:linear} that the $k=0$ mode is always marginally stable, and that the linear instability is controlled only by the combination $Ca^2$ and does not depend individually on the strengths of the smectic alignment and the twist-composition coupling.   At a critical value of  $Ca^2$ determined by $(Ca^2 -2-\lambda_{\psi})^2 = 8\lambda_{\psi}$, the mode with $k_{\textrm{max}} = (2/\lambda_{\psi})^{1/4}$  becomes unstable\cite{supinfo}. Fig.~\ref{fig:dispersion} shows the largest eigenvalue $\omega(k)$ of the linear stability matrix in Eq.(\ref{LSM}) for different parameters.
For any non-zero value of $\lambda_{\psi}$, the instability thus occurs at a finite $k$,  which demonstrates that the instability of the homogeneous phase is to microphase domains of  a finite size.
The transition from a macroscopically phase separated state (infinite domain size, $k=0$) to a microphase separated state should  thus be accompanied by a discontinuity in the domain size \cite{supinfo}.

Numerical analysis of Eqs.(\ref{dyn1}-\ref{dyn2}) verifies this discontinuous change in the domain size.  We solve Eqs.(\ref{dyn1}-\ref{dyn2}) numerically by using an implicit convex splitting scheme to evolve the equation  for $\psi$ and the forward Euler method to evolve the director field \cite{supinfo}.  We initialize the system with random compositional fluctuations around a homogeneous mixture with $\psi=0.2$  and we explore the phase space spanned by  $C$ and $a$. For most of the results shown here, we choose $\lambda_{\psi}=0.1$, as the interface width in the experiments is found to be much smaller than the twist penetration length \cite{sharma2014hierarchical}. Also, we set $q_{0}=0.1$ as the preferred chiral twists of the two species of rods in the experimental system are not equal. The phase diagram obtained from  numerics are shown in Fig. \ref{fig:phase}. It is evident that linear stability analysis captures all qualitative aspects of the numerically determined phase diagram.   The steady state domain sizes obtained from numerics are shown in Fig. \ref{fig:dispersion} and clearly demonstrate the discontinuous change accompanying the phase transition.

The formation of finite size domains is controlled by a competition between chirality and interfacial tension. A similar competition exists even in a chiral membrane of a single species, where the interfacial tension exists between the membrane edge and the bulk polymer suspension. A theoretical analysis of this system \cite{pelcovits2009twist} showed a transition between membranes of finite size and unbounded macroscopic membranes. Within such a membrane, the twist is expelled to the edge, decaying over a length $\lt$, and the membrane size grows continuously as the transition is approached. Here we see that introducing a second species with opposite handedness into such a membrane provides a mechanism for the twist to penetrate the interior of the membrane. As shown in Fig. \ref{fig:phase}, the director twists at the edge of each domain, and then untwists (twists in the opposite direction) into the background. This twist is confined to within approximately $\lt$ of a domain edge. The ability of the interface to accommodate twist is the mechanism that leads to the formation of microdomains in the region of parameter space where each species by itself would form a macroscopic membrane.

\begin{figure}[h!]
\begin{center}
{\raisebox{0.08\height}{\includegraphics[width=0.5\textwidth]{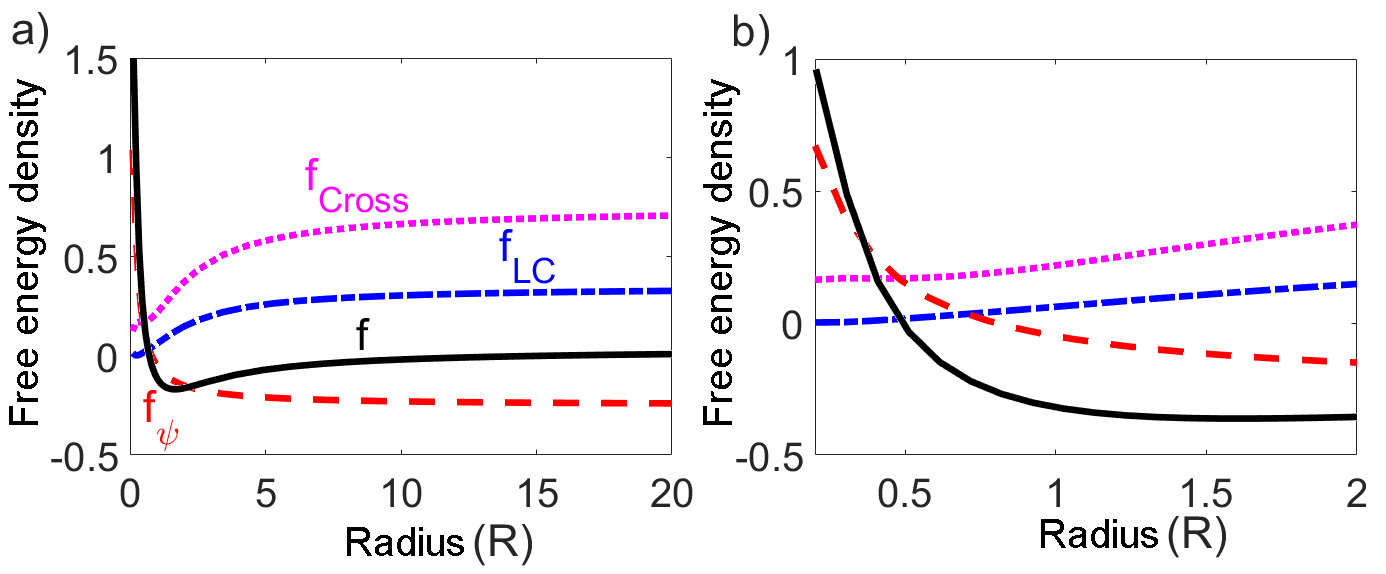}}}
\end{center}
\caption{(color online)  a) The free energy density in a domain is shown as a function of its radius $R$ for $C=30$ and $a=0.7$ calculated using a quasi-static approximation along with the different contributions to it, $f_{\textrm{LC}}$, $f_{\textrm{$\psi$}}$ and $f_{\textrm{Cross}}$. The radius of the droplet is varied from $0.2$ to $20$ in terms of the twist penetration length. b) The same free energy density contributions with a focus on small R.
}
\label{fig:energy}
\end{figure}

 To quantitatively unfold this mechanism, and to understand the discontinuous change in domain size that occurs at the transition to bulk phase separation, we examine how the spatial variations in $\psi$ and $\mathbf{\hat{n}}$  influence the free energy Eq. (\ref{eq:GLequation}). To this end, we calculate the free energy of  a domain of radius $R$ of one species in a background of the other. We do so by assuming profiles for $\psi$ and $\mathbf{\hat{n}}$ that are consistent with the results obtained from numerical integration [\cite{supinfo} section 2]. The optimal domain size is then determined by the value of $R$ at each $C_0$, $a$, $q_0$, $\lambda_\psi$, for which the free energy is minimized (Fig. \ref{fig:energy}). The resulting domain sizes are consistent with those obtained from linear stability analysis and numerical integration (Fig. \ref{fig:dispersion}).

 The origin of discontinuity in domain size is revealed by examining the variations in different contributions to the free energy density ($f_{\textrm{LC}}$, $f_{\textrm{$\psi$}}$ and $f_{\textrm{Cross}}$) as the droplet size changes. Fig. \ref{fig:energy} shows these variations for a parameter set in the microphase separation regime. Note that in an extensive system with clear scale separation between bulk and interface, the interfacial contribution to a free energy density decays with increasing domain size, while the bulk contribution remains constant. In contrast, we see that $f_{\textrm{LC}}$ and $f_\textrm{Cross}$ are super-extensive for small domain sizes, only becoming extensive asymptotically. This superextensivity is significant only for domain sizes of the order of the twist penetration length ($R\sim5 \lambda_\textrm{t}$). Thus, finite-sized domains appear only when the increase in  $f_{\textrm{LC}}$ and $f_\textrm{Cross}$ with $R$ is sufficient to outcompete $f_\textrm{$\psi$}$ at these small domain sizes. As $Ca^2$ decreases, the super-extensive behavior diminishes, forcing the critical domain size (at which  $f_{\textrm{LC}}$ and $f_\textrm{Cross}$ dominate over $f_\textrm{$\psi$}$) to larger $R$. At the threshold value of $Ca^2$,  $f_{\textrm{LC}}$ and $f_\textrm{Cross}$ become extensive before dominating over the interfacial tension, and macrophase separation sets in.
\begin{figure}[h!]
\begin{center}
{\raisebox{0.08\height}{\includegraphics[width=0.5\textwidth]{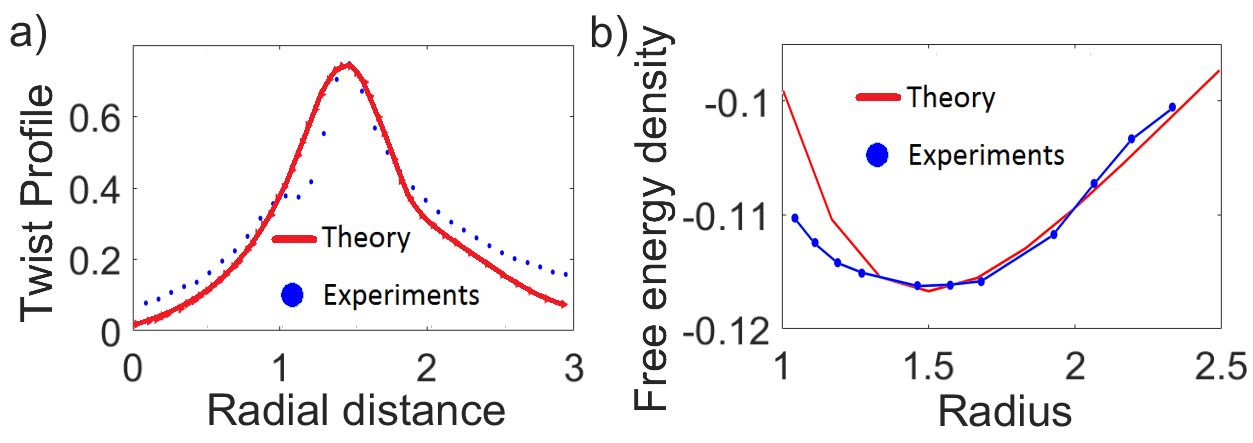}}}
\end{center}
\caption{(color online)  a) The theoretical twist profile obtained through dynamics in our study is compared with the one obtained from the experiments\cite{sharma2014hierarchical}. b) The free energy density profile as a function of radius of droplet is compared between theory and experiments.
}
\label{fig:compare}
\end{figure}

The source of the super-extensive growth in  $f_{\textrm{LC}}$ and $f_\textrm{Cross}$ can be understood from the dependence of twist profiles on $R$. For large $R$, twist decays exponentially from the domain edge (Fig. 2 in \cite{supinfo}); thus ensuring scale separation between the bulk and the interface. On the other hand, such a separation does not exist for small domains where the twist penetrates to the center of the domain.


 In conclusion, we have  presented a theory of microphase separation in membranes, which is driven by chirality of its constituent entities.   The underlying mechanism of microphase separation can be traced to the the gain in twist energy in these structures, which can  accommodate twist at the boundaries of domains.  We have provided quantitative analysis that unfolds the precise factors leading to the appearance of microdomains.   We have also shown that the microdomains have a natural length scale determined by the twist penetration depth, and therefore the domain size does not increase continuously as the system transitions to the macrophase separated state.  Domains that are much larger than the twist penetration depth fail to gain enough free energy from the twisting at the interface to compensate for the free energy cost of creating an interface where the composition changes.  By reducing $\lambda_{\psi}$, this limiting length can be made larger, however the transition is discontinuous for all finite values of $\lambda_{\psi}$.    This feature of the microdomains is appealing from the perspective of creating nanostructures since the domain size can be tightly controlled.

\paragraph{Acknowledgment:} This work was supported by the Brandeis University NSF MRSEC, DMR-1420382. Computational resources were provided by the NSF through XSEDE computing resources (Stampede) and the Brandeis HPCC which is partially supported by  DMR-1420382. We gratefully acknowledge Robert Meyer, Robert Pelcovits, Zvonimir Dogic, and Prerna Sharma for helpful discussions; we also thank Prerna Sharma for providing her experimental data.



\end{document}